# Benchmarking Web-testing - Selenium versus Watir and the Choice of Programming Language and Browser

Miikka Kuutila, M3S, ITEE, University of Oulu, Finland

Mika Mäntylä, M3S, ITEE, University of Oulu, Finland

Päivi Raulamo-Jurvanen, M3S, ITEE, University of Oulu, Finland

Email:
firstname.lastname@oulu.fi,
Postal address:
P.O.Box 8000
FI-90014 University of Oulu

Abstract

Context: Selenium is claimed to be the most popular software test automation tool. Past academic works have mainly neglected testing tools in favour of more methodological topics.

Objective: We investigated the performance of web-testing tools, to provide empirical evidence supporting choices in software test tool selection and configuration.

Method: We used 4*5 factorial design to study 20 different configurations for testing a web-store. We studied 5 programming language bindings (C#, Java, Python, and Ruby for Selenium, while Watir supports Ruby only) and 4 Google Chrome, Internet Explorer, Mozilla Firefox and Opera. Performance was measured with execution time, memory usage, length of the test scripts and stability of the tests.

Results: Considering all measures the best configuration was Selenium with Python language binding for Google Chrome. Selenium with Python bindings was the best option for all browsers. The effect size of the difference between the slowest and fastest configuration was very high (Cohen's d=41.5, 91% increase in execution time). Overall Internet Explorer was the fastest browser while having the worst results in the stability.

Conclusions: We recommend benchmarking tools before adopting them. Weighting of factors, e.g. how much test stability is one willing to sacrifice for faster performance, affects the decision.

Keywords: Software testing, Selenium, Watir, Webdriver, test automation, web-testing

# 1. Introduction

*If it could save a person's life, could you find a way to save ten seconds off the boot time?* -Steve Jobs

Internet reached over 3 billion users in 2014 and the number of users has since grown (InternetLiveStats., 2016). With increased number of users and applications in the web, e.g. Social Media (van Dijck, 2013), Internet of Things (Gubbi et al., 2013) and Cloud based solutions (Gunawi et al., 2014), there comes the growing need for testing these services and applications.

A recent online survey states that Selenium is the most popular software testing tool in the industry (Yehezkel, 2016), a fact reflecting how the popularity of web-based solutions also affects the popularity of the testing tools. Similarly, our recent paper indicates that Selenium is the most popular pure testing tool when using combined criteria consisting of things like: number of survey responses, Google web hits, Twitter tweets and Stackoverflow questions (Raulamo-Jurvanen et al., 2016). Watir also appeared in the responses of those surveys, but based on references by the respondents it was less popular tool than Selenium (Raulamo-Jurvanen et al., 2016; Yehezkel, 2016).

Speed of software development and test automation is highly important in software development and in particular web-development success. This is supported by several reports. According to another industrial survey (Vaitilo and Madsen, 2016) investments in test automation "will be mandatory for coping with the growing demand for velocity". With this demand for velocity, the speed of software testing becomes an important problem. Furthermore, testing is often part of development practice called continuous integration (CI) where the development team integrates their work frequently, and the build is automated along with the tests Fowler (2006). Continuous integration also makes testing continuous. Martin Fowler (2006) sees rapid feedback as one of the primary benefits behind CI, while in his experience testing is the bottleneck in behind increased build times. Therefore, faster performance in testing can lower the build times and enable more rapid feedback, or can allow for more time consuming and comprehensive test sets. Fowler's post is ten years old but support for this notion is found in more recent advice from test automation professionals highlighting the importance of feedback, not only from the CI machine but also from the developers' personal test environments: "Fast feedback loops while you work are incredibly important. In many ways, the length of time to run a single test against my local changes is the biggest predictor of my productivity on a project" (McIver, 2016).

Past work on software testing has mostly focused on more methodological issues in software testing. For example, plenty of academic work has focused on regression test selection and a survey by Yoo and Harman (2012) contains as many as 189 references on this topic. We claim that focusing on methodological issues alone is not enough if we want to do industrially relevant software engineering research. Instead, we should give more emphasizes on studying the tools of our trade. For instance, surprisingly little is known about the performance of web-testing tools that have high penetration among practitioners as we demonstrate in Section 2.

Finally, our modified version of Jobs's quote says "*If it could save a person's life, could you find a way to save ten seconds off the build time when using Selenium?*" Consequently, we could. First, according to IDC report (Hilwa, 2013) there were 18.5 million professional software developers in the world in 2014. If we assume that only 100,000 of those are using Selenium, a safe underestimate, as part of continuous testing where they run the tests 10 times per working day, then 10 second savings for each test run would result in about 80 years saved annually.

Given the importance of speed and popularity of Selenium this paper aims to provide the best evidence to date on the performance of various configurations of Selenium and Watir tools. Such evidence can help in tool and configuration choices made in the industry and help designing more experiments for benchmarking the tools. This paper is organized as follows: Section 2 reviews the existing literature, Section 3 presents our research methods, Section 4 shows the results while Sections 5 and 6 discuss the results and provide the conclusions.

## 2. Literature Review

This section consists of six sub-sections. First, we present the existing literature reviews of web testing. Then we provide a brief introduction to Selenium and Watir in sub sections 2.2 and 2.3. Sub section 2.4 reviews the scientific literature of Selenium and Watir. In sub section 2.5 we review the blogosphere focusing on Selenium and Watir. We do this as the tools are widely adopted in the industry and, thus, grey literature could provide additional, albeit not scientific, evidence. Finally 2.6 summarizes the gaps in the current knowledge.

### 2.1.     Literature reviews of web testing

We found two literature reviews focusing on web testing. A recent survey (Li et al., 2014) looked into techniques for Web application testing, covering techniques such as model-based testing, mutation testing, fuzz testing and random testing. The authors state that different techniques have different roles for testing, some techniques are better for finding faults and some for making sure that the application is adequately tested. The biggest challenges to testing are seen as the ubiquity of the Web and that Web applications support complex transactions in a relatively short period of time and all the while security has to be ensured. With respect to actual tools only a few mentions are made and no benchmark comparison are referred to. For example, it is mentioned that "The most popular AJAX testing tools are currently capture-replay tools such as Selenium, Sahi, and Watir".

Similarly, Dogan, Betin-Can & Garousi (2014) performed systematic literature review into web application testing. Majority of their work focuses on different testing techniques. They report that tools that are related and proposed for Web application testing (WAT) approaches are becoming increasingly available through downloads. They also report that measuring test effort/time was the most frequent in the studies observed, and code coverage was used as a metric most frequently with it. However, they provide no mentions of tools like Selenium or Watir.

With respect to our work, both literature reviews yield to a similar conclusion: There is lack of studies focusing on the efficiency of the popular web testing tools used in the industry.

### 2.2.     Introduction to Selenium

SeleniumHQ (2016a) is a set of open source test automation tools that are primarily used for automating the tests of web applications. Scripts used by Selenium can be coded by hand, created with Selenium IDE (Integrated Development Environment) or recorded with a browser add-on. Selenium IDE produces test scripts that are written in Selenese, a domain specific language just for the use of Selenium. Test scripts coded by hand can be written in a variety of languages, including Java, C#, Perl, PHP, Python and Ruby. Additionally, Selenium Grid allows the distribution of tests to several machines for parallel execution and managing of several testing environments from a central point.

Selenium Webdriver makes direct calls to each browser. These calls are made by browser specific

Webdrivers. The browser specific Webdriver also retrieves the results from the browser it is driving. Bindings to Selenium Webdriver API are language specific, for example when writing test scripts in Java, Java bindings need to be used. Also depending on Webdriver used, some additional measures such as wait statements need to be added to the test scripts.

Using Selenium Webdriver leaves the tester the basic choices of which browser to use, e.g. Firefox or Chrome, and the choice which script level bindings to use.

## 2.3. Introduction to Watir

In Watir's own website (Watir, n.d.) Watir is described to be "an open-source family of Ruby libraries for automating web browsers". The name is an acronym of "Web Application Testing in Ruby" and consequently, the scripts for Watir are written in Ruby only. First version of Watir was Watir Classic (Watir, 2016) and it is much like Selenium RC, that existed prior to Selenium Webdriver. Similarly, Classic version did not use Webdriver, but instead it used Ruby's object linking capabilities to automate Internet Explorer for Windows. Unlike Selenium, Watir does not have an official IDE for recording and editing test scripts. To use Watir Webdriver one needs to write test scripts by hand using Ruby. The architecture of Watir Webdriver does not differ from Selenium Webdriver, as Watir Webdriver is a wrapper of Selenium Webdriver (Zeng, 2014).

Both Watir and Selenium were originally developed without Webdriver, which aims to mimic the behaviour of the user and controls the browser itself. The test scripts made for both are run through Webdriver. There are existing Webdrivers for all the major browsers including Chrome, Internet Explorer, Firefox, Opera and Safari. Webdriver can also be run in headless mode without a GUI (e.g. HtmlUnit and PhantomJS). There have been efforts to make Webdriver an internet standard and working draft has already been released (W3C. et al., 2016).

The main features and facts about both Selenium and Watir are recapped in Table 1. The relationships between different versions and especially the logic behind naming them can be quite confusing and difficult. The reasons for this are explained by Zeng (2014).

**Table 1.** *Summing up main features and facts about Selenium and Watir*

|  | Selenium | Watir |
|---|---|---|
| **Support for Programming languages** | **By SeleniumHQ:** Java, C#, Ruby, Python, JavaScript, Selenese. **By 3rd party:** Perl, PHP, Haskell, Objective-C, R, Dart, Tcl | Ruby **By 3rd party:** Port to .Net in WatiN (support for all 20+ .Net languages like C# and VB.NET), Port to Java in Watij. |
| **Browsers supporting Webdriver** | Firefox, Internet Explorer (6,7,8,9,10 and 11), Safari, Opera, Chrome, HtmlUnit (headless) | Firefox, Internet Explorer, Safari, Opera, Chrome, HtmlUnit (headless) |
| **Members of the family** | Selenium 1.0 (Remote Control), Selenium IDE, Selenium Grid, Selenium 2.0 (Webdriver) | Watir Classic, Watir Webdriver, Watirspec |
| **Original Release** | 2004 | First release found in GitHub 1.6.0 from 2008. |
| **License** | Apache License 2.0 | BSD |

## 2.4.    Selenium and Watir in scientific literature

There have been several studies done with Selenium and Watir. However, the conclusions have often been based on opinions or limited in benchmarking results as summarized in Table 2. Additionally, the existing benchmarking results seemed to conflict. Thus, we felt an additional study was merited for these highly popular web-testing tools. The remaining of this section provides further details on the prior works.

**Table 2.** *What are Conclusions based on?*

| Study | Conclusion based on: |
|---|---|
| **Javvaji, Sathiyaseelan & Selvan (2011)** | Opinion based on the usage of Selenium. |
| **Singh and Tarika (2014)** | Measured execution time of logging into gmail account. Grading and comparing different features of Selenium, Watir and Sikuli tools. |
| **Gogna (2014)** | Opinion on the use of Selenium and Watir. |
| **Angmo and Sharma (2014)** | Performance measurement in execution speed (in milliseconds) of how many windows the software can open. Grading of different features of Watir and Selenium tools by comparing them to each other. |
| **Li, Das & Dowe (2014)** | Literature review. |
| **Leotta, Clerissi, Ricci & Spadaro (2013)** | Measurement of time (in minutes) over lines of code modified. |
| **Kongsli (2007)** | Experiences on the usage of Selenium for security testing. |
| **De Castro, Macedo, Collins & Dias-Nieto (2013)** | Execution times and experiences of a self-made tool that is an extension to Selenium tool family. |

Javvaji, Sathiyaseelan & Selvan (2011) give introduction to the different qualities and technical overview of Selenium, and some specifics on the use of Selenium related to data driven testing. The authors tell about their experiences in the form of "Do's, Don'ts, Merits & Dependencies". Overall, we found this paper to be completely opinion based.

Singh and Tarika (2014) offer a comparative look at three different open source automation testing tools, with Selenium, Watir, and Sikuli. The tools were compared on seven different factors: recording capabilities, execution speed, scripts generation, data driven testing, ease of learning, testing reports/output and supplement features. For each of these factors, a score was given in the scale of 1 to 5 and the mean of these grades was used to rank these tools. In the conclusions, some interesting reflection was given to the scores. Selenium was ranked as the best one of the three tools, thanks to its recording features, ease of learning, features that support data driven testing and for the general support of 3rd party application integration. Watir was ranked second in the study, outclassing Selenium in execution time but falling short of Selenium in other categories. The authors thought that Watir was mainly lacking in programming languages for test cases, importing the scripts and native support of recording test cases (instead of scripting them by hand). Sikuli was scored the best on test execution speed and ease of learning, but came short on recording capabilities and supplement features. Another weakness lowering Sikuli's score was that it only supports scripting in Python.

Gogna (2014) introduces the basic features of both Selenium and Watir, but does not venture into comparing them at length. On conclusions, the author sees the fact of Watir using Ruby for scripts as a strong suit compared with Selenium's vendorscript (Selenese). The author also sees trouble on recording Iframes, frames and popup windows with Selenium, but these can be accessed using API with Watir. Deep learning curve when switching from Selenium IDE to Selenium RC is also noted. Though there is no clear statement in favor of either tools, a reader gets a sense of preference for Watir by the author.

Angmo and Sharma (2014) evaluated the performance of both Selenium and Watir, and they used multiple versions of tools from the Selenium family. For Selenium, they used IDE, Remote Control, Webdriver and Grid, while for Watir they used only Watir Webdriver. Selenium IDE comes out on top for execution speed while Selenium Webdriver beats Watir Webdriver.

In the systematic literature review by Li et al. (2014) the manual construction of test cases is seen as the biggest limitation to Watir and Selenium, and a random testing framework Artemis is suggested to be used together with them. This fact is intensified because JavaScript (which is used on most web pages) is dynamic and event driven, meaning that depending on what is tested it may be difficult to ensure "a particular path" through the software being tested.

Case study by Leotta, Clerissi, Ricci & Spadaro (2013) found out that repairing test cases was more efficient using ID locators than path locators for test suites using Selenium Webdriver. The difference was seen on both time to repair and lines of code modified. Reducing the maintenance of test suites can reduce maintenance costs related to test automation significantly.

There has been a number of papers and studies that propose additional ways of using Selenium for testing, such as de Castro, Macedo, Collins & Dias-Nieto (2013) and Kongsli (2007). While they may not always be relevant to show the effectiveness of Selenium for automation testing specifically, they may imply good modifiability for usage. Kongsli (Kongsli, 2007) showed that Selenium can be also used in testing the security of web applications. The biggest limitation of Selenium in this context was seen to be that some vulnerabilities could be very hard to expose through the web interface, and as the core of Selenium is JavaScript it is subject to the restrictions of JavaScript when running a browser. De Castro et al. (2013) presented an extension that allows the performing of tests on web applications with databases. This extension is called SeleniumDB and allows testing for applications using MySQL and PostgreSQL databases.

Overall, there are no major findings that can be supported by multiple sources, quite the opposite in fact. Singh and Tarika (2014), for example, find that Watir Webdriver has a faster execution time than Selenium IDE plugin for Firefox, but Angmo and Sharma (2014) claim that Selenium Webdriver is faster than Watir Webdriver when it comes to performance. Both of the studies use only one set of instructions to measure execution speed (logging into Gmail account vs. opening instances of a browser), so the results are hard to generalize. Another example of opposite conclusions is that Angmo and Sharma (2014) grade Selenium easier to use than Watir, but Gogna (2014) has the opposite opinion based on her use of the tools.

## 2.5.    Selenium and Watir in blogosphere

Both Selenium and Watir have lots of blogs dedicated to their purposes and the open source communities behind them have collected those together at Watir Blogs (Watir, n.d.) and No Automated Testing (2016). To gather additional evidence we studied both sources. In general, there are lots of blogs that have "how to" content (in particular for Selenium) and those are easy to find in either Watir Blogs (Watir, n.d.) or No Automated Testing (2016). Although, we could not find any properly reported empirical studies in the blogs, we did find experiences. Table 3 recaps the blogs and blog posts introduced by covering the key claims and contents.

**Table 3.** *Recapping the blogosphere*

| Blog | Key Claim or Information |
|------|------------------------|
| **SauceLabs (2009)** | Information about testing methods and practical use of Selenium. |
| **Badle (2010)** | Blog mainly on the development and use of Selenium plugins for Firefox. |
| **Zeng (2014)** | Blog post explaining the history and growth of Selenium and Watir open source projects. |
| **Robbins(2011)** | Review of Watir Webdriver. |
| **Phillips (2012)** | Claim of Watir Webdriver being faster than Selenium Webdriver, gimmicks for the use of Watir and an account of migrating unit tests from Selenium to Watir. |
| **Scott (2010); Scott (2011); Scott (2014); Scott (2016)** | Blog of former Watir developer. Information on the use of Watir and ponderings on the future of Watir. Claim of Selenium being more verbose and less readable than Watir (with an example). Claim of Internet Explorer being a "very non-testable browser". |
| **Hendrickson (2010)** | Report on investigating the programming skills needed in jobs for testing automation. Selenium mentioned in more job ads than Watir. |

## 2.6.    Gaps in current understanding

Angmo and Sharma (2014) and Singh and Tarika (2014) report differing results on test execution speed for Selenium and Watir. Differing results can be explained by comparing different versions of software, i.e. Selenium RC vs Watir Webdriver in Angmo and Sharma (2014) and Selenium Webdriver vs Watir Webdriver in Singh and Tarika (2014). In both of these papers the tasks used for comparing execution time of tests are not taken from "real world", but instead one simple task is used in both of these papers (logging into Gmail account (Singh and Tarika, 2014) and counting the number of opened browser windows in a second (Angmo and Sharma, 2014)).

The studies do not compare different testing configurations, thus leaving questions if the usage of different language bindings or web browsers affects time of execution. Furthermore, usage of memory by the tools is not reported, which could specifically influence the demand for resources when executing the tests on a virtual environment. Additionally, there is no information given regarding the test execution problems or maintainability of the scripts. Our research goal was formed to fill the gaps in existing knowledge.

## 3.  Methodology

This chapter consists of describing the design and execution of the research concluded for this paper. Areas of research were identified by the literature review conducted in Chapter 2.

## 3.1.    Research goal and measures

Our high level research goal is *to find the most performance efficient configuration for web-testing*. We have four measures of performance*: execution time, memory usage, length of test scripts and stability of the tests*. First two measures are typical performance measures. We consider the length of test scripts as an indication of test script maintainability, i.e. the shorter the script the less effort it is required to maintain it. We are aware that it would be better to measure hours spent in the test maintenance, unfortunately, such measure was not available to use. Furthermore, recent evidence suggests that length of code could be considered as a primary surrogate maintenance measure as it appears to be more accurate than for example code smells (Sjøberg et al., 2013). Kamei et al. (2010) suggests this measure is applicable to test code as well. Finally, we consider the stability of the tests on how many test runs fail or terminate abnormally when they should not. These are often referred to as flaky tests. If a particular configuration has many flaky executions, it increases cost in unnecessary test analysis effort.

Our design is 5X4 between subject factorial design (Wohlin et al. 2012, p.98), see table Table 4. We collected all of these measures for 20 (5*4) different configurations. In more detail, we had five different bindings or tools (Watir tool and C#, Java, Python and Ruby bindings for Selenium tool). All of the tools and bindings were executed in four different browsers (Firefox, Chrome, Opera and Internet Explorer). Complete configurations along with unit testing frameworks and other information are introduced next in Table 5. Table 6 contains additional information considering the configurations using C#, specifically the programming environment and the test runner used. Finally, Table 7 shows the hardware we used to execute our tests.

**Table 4.** *5x4 Design*

|  | Browsers | | | |
|---|---|---|---|---|
| **Library/Bindings** | Chrome | Firefox | Internet Explorer | Opera |
| C# | C#/Chrome | C#/Firefox | C#/IE | C#/Opera |
| Java | Java/Chrome | Java/Firefox | Java/IE | Java/Opera |
| Python | Python/Chrome | Python/Firefox | Python/IE | Python/Opera |
| Selenium/Ruby | Ruby/Chrome | Ruby/Firefox | Ruby/IE | Ruby/Opera |
| Watir/Ruby | Watir/Chrome | Watir/Firefox | Watir/IE | Watir/Opera |

**Table 5.** *Every configuration used*

| Programming language | Tool | Browser | Webdriver | Unit testing framework |
|---|---|---|---|---|
| **C#** | Selenium 2.48.0.0 | Firefox 42.0 | Webdriver with installation of Selenium 2.48.0 for C# | Microsoft.VisualStudio. QualityTools. UnitTestFramework Version 10.0.0.0 |
| **C#** | Selenium 2.48.0.0 | Chrome 46.0.2490.80 | ChromeDriver 2.19.346078 | Microsoft.VisualStudio. QualityTools. UnitTestFramework Version 10.0.0.0 |
| **C#** | Selenium 2.48.0.0 | Internet Explorer 11.0.96000.18098 | InternetExplorerDriver server (32-bit) 2.4.8.0.0 | Microsoft.VisualStudio. QualityTools. UnitTestFramework Version 10.0.0.0 |
| **C#** | Selenium 2.48.0.0 | Opera 30.0.1835.125 | OperaChromiumDriver 0.1.0 | Microsoft.VisualStudio. QualityTools. UnitTestFramework Version 10.0.0.0 |
| **Java JDK 1.8.0_60** | Selenium 2.47.1 | Firefox 42.0 | Webdriver with installation of Selenium 2.47.1 for Java | JUnit 4.12 |
| **Java JDK 1.8.0_60** | Selenium 2.47.1 | Chrome 46.0.2490.80 | ChromeDriver 2.19.346078 | JUnit 4.12 |
| **Java JDK 1.8.0_60** | Selenium 2.47.1 | Internet Explorer 11.0.96000.18098 | InternetExplorerDriver server (32-bit) 2.4.8.0.0 | JUnit 4.12 |
| **Java** | Selenium | Opera | OperaChromiumDriver 0.1.0 | JUnit 4.12 |

| JDK 1.8.0_60 | 2.47.1 | 30.0.1835.125 | | |
|---|---|---|---|---|
| **Python 2.7.10** | Selenium 2.46.0 | Firefox 42.0 | Webdriver with installation of Selenium 2.46.0.0 for Python | Unittest with installation of Python 2.7.10 |
| **Python 2.7.10** | Selenium 2.46.0 | Chrome 46.0.2490.80 | ChromeDriver 2.19.346078 | Unittest with installation of Python 2.7.10 |
| **Python 2.7.10** | Selenium 2.46.0 | Internet Explorer 11.0.96000.18098 | InternetExplorerDriver server (32-bit) 2.4.8.0.0 | Unittest with installation of Python 2.7.10 |
| **Python 2.7.10** | Selenium 2.46.0 | Opera 30.0.1835.125 | OperaChromiumDriver 0.1.0 | Unittest with installation of Python 2.7.10 |
| **Ruby 2.1.7p400** | Selenium 2.47.1 | Firefox 42.0 | Webdriver with installation of Selenium 2.47.0 for Ruby | Test-unit 2.1.7.0 |
| **Ruby 2.1.7p400** | Selenium 2.47.1 | Chrome 46.0.2490.80 | ChromeDriver 2.19.346078 | Test-unit 2.1.7.0 |
| **Ruby 2.1.7p400** | Selenium 2.47.1 | Internet Explorer 11.0.96000.18098 | InternetExplorerDriver server (32-bit) 2.4.8.0.0 | Test-unit 2.1.7.0 |
| **Ruby 2.1.7p400** | Selenium 2.47.1 | Opera 30.0.1835.125 | OperaChromiumDriver 0.1.0 | Test-unit 2.1.7.0 |
| **Ruby 2.1.7p400** | Watir 0.9.0 | Firefox 42.0 | Webdriver with installation of Watir 0.9.0 | Test-unit 2.1.7.0 |
| **Ruby 2.1.7p400** | Watir 0.9.0 | Chrome 46.0.2490.80 | ChromeDriver 2.19.346078 | Test-unit 2.1.7.0 |
| **Ruby 2.1.7p400** | Watir 0.9.0 | Internet Explorer 11.0.96000.18098 | InternetExplorerDriver server (32-bit) 2.4.8.0.0 | Test-unit 2.1.7.0 |
| **Ruby 2.1.7p400** | Watir 0.9.0 | Opera 30.0.1835.125 | OperaChromiumDriver 0.1.0 | Test-unit 2.1.7.0 |

**Table 6.** *Complete information for C# development environment and test runner user.*

| **Visual studio & C# version** | **Microsoft Visual Studio Community 2015** **Version 14.0.23107.0** **Microsoft .NET Framework** **Version 4.6.00081** **Installed Version: Community** |
|---|---|
| **Test runner needed for command line execution** | Microsoft Test Execution Command Line Tool Version 14.0.23107.0 (MS Test) |

**Table 7.** *System information for System under Test.*

| **Part of System** | |
|---|---|
| **Operating system** | Windows 8.1 professional 64-bit |
| **CPU** | Intel Xeon E3-1231 v3 3,4GHz, 8 cores |
| **Random Access Memory** | 2x 4096MB DDR3-1600 CL9 in Dual Channel |
| **GPU** | 2x NVIDIA GeForce GTX 970 4GB in SLI configuration |
| **Storage** | Kingston SH103S3210G SSD |
| **Motherboard** | Z97 chipset |

The whole system uses about 1340 megabytes of virtual memory (or 1,34 gigabytes) after start-up. This was measured several times after start-ups using Windows Performance Monitor, introduced in Section 3.5.

## 3.2.    System under test

Mozilla (2016) is a website for downloading and distributing add-ons to Mozilla's Firefox browser. It contains dynamic menus, ajax content and can be described to generally resemble online stores. Tests implemented for the research system contain simple tests such as tests for making sure a specific web element is working as intended and, also for more complex actions such as navigating dynamic menus, logging in to an account and posting a review for an add-on.

## 3.3.    Test set

To answer the research questions, we created a test set that contained equivalent implementations for our five different bindings or tools (Watir tool and C#, Java, Python and Ruby bindings for Selenium tool). To increase validity in our study, we searched for an existing test set that would not only be publicly available but would also be used in the real world. We ended up selecting tests for Mozilla's add-ons site (Mozilla, 2016) that is used for downloading and installing plugins for Firefox browser. In the aftermath, we do realize that our selection of the site could favour Firefox browser. However, there was lack of repositories to choose from at the time of making that decision. The selected repository contained wide variety of tests, its dependencies to 3rd party frameworks and extra code could be removed in a relatively timely manner, and lastly execution of the tests required no extra effort in hosting specific parts of systems. From all the tests, we selected 20 tests at random to our test set. Tests were not chosen from two files: test_installs.py and test_api_only.py. File test_installs.py was not used because these tests would have worked only using Firefox browser. File test_api_only.py was not used because it contained only tests that did not use Selenium Webdriver.

The original tests introduced in Table 8 used Selenium Webdriver with Python bindings, but even the original tests had to be re-implemented for the purpose of this research. This was because they used 3rd party frameworks and libraries for test flagging and assertions, which were not available as such for other programming language bindings. The wrapper used to control and conceal HTML element selectors was also in the original implementation, which was not used in the re-implementation for this research. This made code comparison easier.

Test set constructed for every configuration contained these 20 randomly chosen tests developed applying a basic unit testing framework (which came with the installation of each programming environment, see Table 5), with setup and teardown methods. Both setup and teardown methods were run between every individual test, setup initializing the browser and teardown closing it.

**Table 8.** *Description of the tests contained in set used for measuring execution speed and memory usage.*

| Test name on original file | Asserts | Short description |
|---|---|---|
| **test_that_featured_themes_exist_on_t he_home** | 1 | Asserts that featured title menu on main page contains the text "Featured Extensions, see all". |
| **test_addons_author_link** | 2 | Compares main page's add-on author name to the one in add-on's own page. |
| **test_that_check_if_the_extensions_are _sorted_by_most_user** | 2 | Goes to featured extensions page and presses button to sort extensions by most users. Makes sure extensions are sorted. |
| **test_that_checks_if_subscribe_link_ex ists** | 1 | Checks that feature extensions page has the subscribe button. |
| **test_featured_tab_is_highlighted_by_ default** | 1 | Checks that after clicking the featured collections link in the main page, the collections are sorted by features (and thus featured is highlighted). |
| **test_create_and_delete_collection** | 3 | Test logs into account, makes an empty collection with uuid description. Checks description and deletes previously created collection. (3 assert statements, one in a loop.) |
| **test_that_clicking_the_amo_logo_load s_home_page** | 3 | Loads main page, clicks the amo logo and checks that the main page is still displayed. |
| **test_that_other_applications_link_has _tooltip** | 1 | Gets other applications html element from the main page, checks that it has the right kind of tooltip. |
| **test_the_search_box_exist** | 1 | Goes to main page and sees that html element with the ID "search-q" exists, checks it's displayed. |
| **test_that_new_review_is_saved** | 4 | Makes a review for add-on firebug and puts a timestamp on the review. Goes to users own page to check the info on saved review. |
| **test_that_searching_for_cool_returns_ results_with_cool_in_their_name_des cription** | 3 | Inserts cool on the search box on main page and checks that all the results on next page contain cool on their name or description. (3 asserts total, 2 inside try-catch.) |
| **test_sorting_by_newest** | 2 | Searches with term "firebug", sorts the results with "newest". Checks that the dates of shown add-ons are newest first. |
| **test_that_searching_for_a_tag_return s_results** | 2 | Searches with term "development" and searches with it as a tag. Checks that both searches yield results. |
| **test_that_verifies_the_url_of_the_stati stics_page** | 1 | Goes to the "firebug" add-ons statistics page, verifies its url. |
| **test_the_recently_added_section** | 3 | Goes to themes menu from main page, checks that recently added themes are sorted by date. |
| **test_that_most_popular_link_is_defau lt** | 2 | Checks that most popular option on themes menu is either bolded or its font weight is over 400 meaning its selected. |
| **test_that_external_link_leads_to_add on_website** | 2 | Goes to "memchaser" add-ons page and clicks it external homepage link. Makes sure the link is the same you are directed to. |
| **test_user_can_access_the_edit_profile _page** | 7 | Logs user in, checks that edit profile page is accessible. |
| **test_that_make_contribution_button_ is_clickable_while_user_is_logged_in** | 3 | Goes to firebug add-ons page, logs user in, clicks contribute button and checks confirm button is clickable. |
| **test_the_logout_link_for_logged_in_u sers** | 4 | Logs user in and checks login status. Logs user out and checks login status. |
| | **48** | **Total assert statements.** |

## 3.4.     Developing tests

To make our approach transparent we provide complete access to our measurements and the source code at http:goo.gl/gEqX07. In addition the site contains extra material on the implementation issues and solutions that were experienced by the first author.

## 3.5.     Test set execution and measurement in system under test

The test set was constructed for every configuration, 20 in all and it was run using Windows PowerShell Version 4.0 (Build 6.3.9600.42000). This was to ensure the same way of getting the execution time of the test set for every time it was run. The test set was run by using command: "Measure-Command{"execute command" | Out-Default}". For example running test set using python bindings and Firefox browser, the command looked something like: "Measure-Command{python .\firefoxpython.py | Out-Default}". Because Measure-Command command usually only gives out execution time, but not any of the other information, out-default option was needed to be added to see what the programs print on commandline. This was e*specially to ensure that measurements* were not taken from runs where one or more errors or faults occurred. Java projects were compiled as runnable jar files and run by measure-command, for example: "Measure-Command{java -jar firefoxjava.jar | Out-Default}". For C# configurations MSTest runner was invoked, for example: "Measure-Command{MSTest /testcontainer:firefoxC.dll | Out-Default}".

Measuring memory usage while the tests were run was done by using Windows Performance Monitor. Measurement taken during test set execution was the mean of virtual memory used for the whole system (recorded every time the test set was run for each configuration). Recording the use of random access memory was deemed troublesome, because windows memory management tries to maximize its use.

The whole process of executing test sets,measuring execution time and mean of committed bytes in memory was specified as follows:

*1. The computer is started (conditions to run the tests the same for all test sets)*

*2. Windows Performance Monitor is started*

*3. Windows Powershell is started*

*4. The user defined data collector from Windows Performance Monitor is started, for measuring memory usage*

*5. First test set execution is started from PowerShell*

*6. When test set execution finishes successfully, the data collector is stopped thus it making a readable report. If execution finishes with failures or errors, measures are not counted and test set execution is restarted along with the data collector. This means going back to step 4.*

*7. Execution time is read from PowerShell and recorded to a table on another computer.*

*8. Mean of Committed Bytes is recorded from the report and recorded to table on another computer*

*9. Steps four through eight are repeated until ten measurements for both execution time and mean of committed bytes are acquired successfully. The measurements obtained were not sequential, meaning sometimes the test set was run 10 or even 15 times to obtain the measurements needed because of false positives errors.*

All in all, 30 measurements for both execution time and mean of virtual memory used were taken for each configuration. Measurements were taken at series of ten, to diminish the influence of latency and level of responsiveness from the server, which could alter due to time of day and level of traffic to the server. The whole design of the research is summed in Table 9. All measurements taken are available at https:goo.gl/gEqX07 .

*Table 9. Research design summed up*

| | |
|---|---|
| **Number of test sets** | **1** |
| **Number of test cases in test set** | 20 |
| **Number of configurations test set is tested with** | 20 |
| **Number of measurements taken for each configuration** | 30 |
| **Number of times the test set is run successfully** | 20 * 30 = **600** times |

## 3.6.    Determining if the results follow normal distribution

Before analyzing any tests for normality, blatant outliers that were deemed as measuring errors, were removed from the data sets by trimming. This was done with results that were clearly more than three standard deviations away from the mean. This meant removing 4 results for execution time and 5 results for mean of memory during execution as outliers due to measuring error.

Tests known as Shapiro-Wilk test and Kolmogorov-Smirnov test are tests that can be used for analysing if a sample follows normal distribution. They were introduced by Shapiro and Wilk (1965) and by Kolmogorov (1933) and Smirnov (1948). For testing if the acquired samples followed normal distribution, results of histograms and Shapiro-Wilk test were examined. Shapiro-Wilk test was primarily considered over Kolmogorov-Smirnov test, because every sample contained only 29 to 30 measurements. Sample was considered not to follow normal distribution when p = 0,05. Table 10 shows the results of Shapiro-Wilk test for both execution time in seconds and mean of used virtual memory in bytes. Table 11 shows the same results for Kolmogorov-Smirnov test.

**Table 10.** *Results of Shapiro-Wilk test*

| Measurement | Number of samples which followed normal distribution | Number of samples which did not follow normal distribution |
|---|---|---|
| **Execution time** | 12 | 8 |
| **Virtual memory** | 9 | 11 |

**Table 11.** *Results of Kolmogorov-Smirnov test*

| Measurement | Number of samples which followed normal distribution | Number of samples which did not follow normal distribution |
|---|---|---|
| **Execution time** | 15 | 5 |
| **Virtual memory** | 12 | 8 |

### 3.7.   Used statistical methods

Because of both analysis of Shapiro-Wilk and Kolmogorov-Smirnov tests, the data was presumed to not to follow normal distribution. Because of previously introduced reasons to support the non-normal distribution of data gathered, Wilcoxon signed-rank test introduced by Wilcoxon (1945) was chosen to compare both execution time and memory usage between pairs of configurations. To check the difference between-subject-effects of the browser, the language bindings, and the interaction of the two we used ANOVA and Partial Eta squared effect size measures. Friedman's test would have been a better alternative, however, according to our best knowledge there is no effect size test for Friedman's test, thus, we were forced to use ANOVA even when we could not be sure of our distribution. However, according to statistical guidelines [1] ANOVA "tolerates violations to its normality assumption rather well."

Usage of memory was further analysed for correlation with the order the measurements were taken, to see if there were any tendencies for the usage of memory to grow towards the end. One could assume that memory usage would be higher towards the end of 10 test set execution run. This was tested by investigating Spearman's rank correlation coefficient introduced by Spearman (Spearman, 1904). As the total samples were taken in series of ten, the correlation was investigated for these series. Seven of the total of sixty samples showed statistically significant positive correlation according to Spearman's rho. Thus, it appeared that there was no evidence for the increase in memory consumption towards the end of the test run.

Correlation of execution speed and memory usage between samples of the same configuration were also investigated with both Spearman's rho (Spearman, 1904) and Kendall rank correlation coefficient (Kendall, 1938). Out of twenty comparisons, both statistics identified two correlations, one positive and one negative. This was interpreted as execution time and memory usage having no correlation.'

## 4.   Results

This section presents the measurements: execution time, memory usage, length of test scripts and stability of the tests.

### 4.1.   Execution speed

Between subject effects was investigated with ANOVA, as explained in Section 3.6. Table 12, shows that Browser, Language bindings, and their interaction have very large effect sizes with Partial Eta Squared being 0.982, 0.989 and 0.964 respectively. Note partial Eta larger than 0.26 is considered as high effect size. In practice this means that nearly all of the variation in the measurement results is explained by these three factors.

[1] https://statistics.laerd.com/statistical-guides/one-way-anova-statistical-guide-3.php

**Table 12.** *Test of Between-Subjects effects with dependent variable of execution time.*

| Source | Type III Sum | df | Mean Square | F | Sig. | Partial Eta Squared |
|---|---|---|---|---|---|---|
| **Corrected Model** | 1440856,73[2] | 19 | 75834,565 | 5141,473 | 0.000 | 0.994 |
| **Intercept** | 30049709,48 | 1 | 30049709,48 | 2037326,468 | 0.000 | 1.000 |
| **Browser** | 460927,802 | 3 | 153642,601 | 10416,744 | 0.000 | 0.982 |
| **Tool** | 756606,391 | 4 | 189151,598 | 12824,202 | 0.000 | 0.989 |
| **Browser * Tool** | 225302,281 | 12 | 18775,190 | 1272,931 | 0.000 | 0.964 |
| **Error** | 8495,758 | 576 | 14,750 | | | |
| **Total** | 31510175,77 | 596 | | | | |
| **Corrected Total** | 1449352,486 | 595 | | | | |

Next, Wilcoxon signed rank-test was used to compare the execution speeds of all the configurations in pairs to find out which configurations performed better than others. For each pair, the configuration that had statistically faster time of execution was given a "win" over its counterpart. If the difference was not statistically significant, the comparison resulted in a draw. This is a typical comparison done for example in machine learning benchmarks, e.g. Arcelli Fontana et al. (2015). Based on these wins a winning percentage was calculated for every configuration. Table 13 lists the winning percentages, execution speeds and effect sizes for different configurations. Overall the results for execution time have low standard deviations, meaning data acquired should be reliable and the way of measuring execution times robust. Effect size measures are used to quantify the size of difference in two groups. Table 13 shows effect size measures in comparison to the winner only. We report Cliff's delta (Cliff, 1993), Cohen's d (Cohen, 1988), and percentage behind the first rank as effect sizes measures. Cliff's delta does not require any assumptions about the shape or deviation of two distributions, i.e. a non-parametric method. As Cliff's delta values quickly reached maximum value (1.0) in Table 13 we also report Cohen's d that is a parametric effect size measure. Finally, the measure percentage behind the first rank is used to highlight the practical difference for our industrial readership that might not be familiar with the two previous measures. Romano, Kromrey, Coraggio & Skowronek (2006) provide thresholds for Cliff's delta values based on Cohen's [1] interpretations of the effect size index d. Lesser delta value than 0.147 is "neglible", lesser than 0.33 is "small", lesser than 0.474 is "medium" and otherwise the value is "large". For Cohen'd reference values are 0.2<="small" effect, around 0.5<= "medium", and 0.8<= is "large". Figure 1 shows box plots highlighting the deviations for different samples.

Table 14 ranks tools/bindings for every browser based on execution time. It shows that C# was the fastest language for Selenium in three out of the four browsers while Watir (that only supports Ruby) was the slowest for all browsers. Overall, results in Table 13 shows that C# Selenium bindings in IE were the fastest configuration while Watir tool in Firefox was the slowest. Finally, we need to note that due to small standard deviations our effect sizes are large even between the fastest and second fastest configuration (Cliff's delta 0.833 and Cohen's d 1.958). However, the practical performance measured as the percentage behind shows only the difference of 3.69% between the first and the second configuration. The median percentage behind the first rank was 30.4% while the measure varied between 3.69% to 91.36%.

---

[2] R Squared 0.994 and adjusted R Squared 0.994

Table 15 ranks browsers for tools/bindings based on execution time and it shows that all configurations using Internet Explorer were the fastest for that particular tool or language binding. There was less agreement on the slowest browser, as Firefox, Opera and Chrome each were the slowest depending on the binding or tool used.

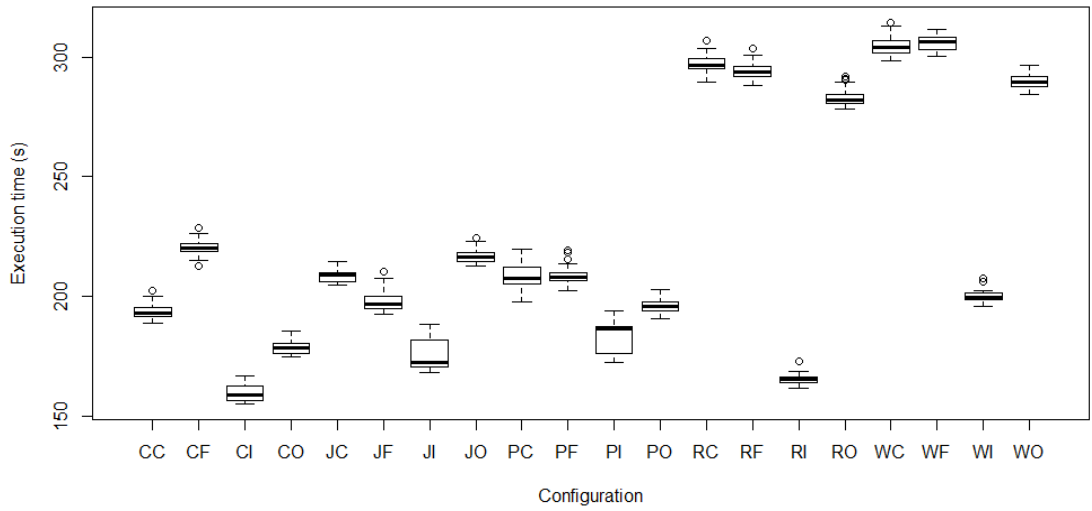

(a)

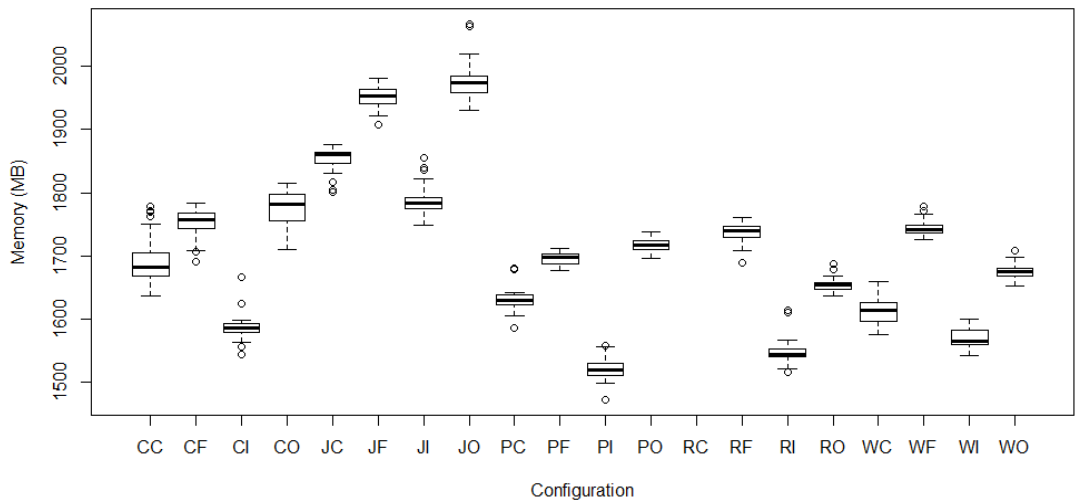

(b)

**Figure 1**. Box plots for samples of execution time and memory usage.

**Table 13.** *Mean, median, standard deviation, effect sizes and number of accepted results for execution time for every configuration. Sorted by winning percentage and mean of execution time.*

| Rank | Configuration | Win % | Mean | Median | St. Dev. | Effect measures in comparison to winner | | | n |
|------|---------------|-------|------|--------|----------|-------------|----------|-----------------|----|
| | | | | | | Cliff's delta | Cohen's d | % behind 1st rank | |
| **1** | C#/IE | 100 | 159.89 | 158.98 | 3.393 | - | - | - | 30 |
| **2** | Ruby/IE | 94.7 | 165.79 | 165.52 | 2.580 | 0.833 | 1.958 | 3.69 | 30 |
| **3** | Java/IE | 89.5 | 175.92 | 172.57 | 6.373 | 1.000 | 3.140 | 10.03 | 30 |
| **4** | C#/Opera | 84.2 | 178.83 | 178.89 | 2.530 | 1.000 | 6.329 | 11.85 | 29 |
| **5** | Python/IE | 78.9 | 184.18 | 186.62 | 6.129 | 1.000 | 4.904 | 15.19 | 30 |
| **6** | C#/Chrome | 73.7 | 193.93 | 193.11 | 3.340 | 1.000 | 10.111 | 21.29 | 29 |
| **7** | Python/Opera | 63.2 | 196.36 | 196.00 | 3.062 | 1.000 | 11.285 | 22.81 | 30 |
| **7** | Java/Firefox | 63.2 | 198.19 | 197.13 | 4.468 | 1.000 | 9.654 | 23.95 | 30 |
| **9** | Watir/IE | 47.4 | 200.04 | 199.68 | 2.698 | 1.000 | 13.098 | 25.11 | 30 |
| **10** | Python/Chrome | 42.1 | 208.52 | 207.66 | 5.094 | 1.000 | 11.236 | 30.41 | 30 |
| **10** | Java/Chrome | 42.1 | 208.63 | 208.91 | 2.772 | 1.000 | 15.732 | 30.48 | 29 |
| **10** | Python/Firefox | 42.1 | 208.99 | 208.22 | 3.960 | 1.000 | 13.316 | 30.71 | 30 |
| **13** | Java/Opera | 36.8 | 217.13 | 216.70 | 3.034 | 1.000 | 17.785 | 35.80 | 30 |
| **14** | C#/Firefox | 31.6 | 220.44 | 220.38 | 3.478 | 1.000 | 17.623 | 37.87 | 30 |
| **15** | Ruby/Opera | 26.3 | 283.23 | 282.06 | 3.673 | 1.000 | 34.553 | 77.14 | 30 |
| **16** | Watir/Opera | 21.1 | 289.77 | 289.28 | 2.988 | 1.000 | 40.627 | 81.23 | 30 |
| **17** | Ruby/Firefox | 15.8 | 293.87 | 293.51 | 3.587 | 1.000 | 38.375 | 83.80 | 30 |
| **18** | Ruby/Chrome | 10.5 | 297.15 | 296.68 | 3.724 | 1.000 | 38.531 | 85.85 | 30 |
| **19** | Watir/Chrome | 0 | 304.42 | 304.05 | 3.742 | 1.000 | 40.465 | 90.39 | 29 |
| **19** | Watir/Firefox | 0 | 305.97 | 306.29 | 3.163 | 1.000 | 44.536 | 91.36 | 30 |

**Table 14.** *Tools/bindings ranked for browsers by their mean of execution time.*

| Browser | 1st | 2nd | 3rd | 4th | 5th |
|---------|-----|-----|-----|-----|-----|
| **Firefox** | Java | Python | C# | Ruby | Watir |
| **Chrome** | C# | Python | Java | Ruby | Watir |
| **Opera** | C# | Python | Java | Ruby | Watir |
| **Internet Explorer** | C# | Ruby | Java | Python | Watir |

**Table 15.** *Browsers ranked for bindings/tools by their mean of execution time.*

| Tool/Binding | 1st | 2nd | 3rd | 4th |
|--------------|-----|-----|-----|-----|
| **C#** | Internet Explorer | Opera | Chrome | Firefox |
| **Java** | Internet Explorer | Firefox | Chrome | Opera |
| **Python** | Internet Explorer | Opera | Chrome | Firefox |
| **Ruby** | Internet Explorer | Opera | Firefox | Chrome |
| **Watir** | Internet Explorer | Opera | Chrome | Firefox |

## 4.2.    Virtual Memory Usage

Results of virtual memory usage were analysed in the same way as the execution speeds the in previous section. Table 16 shows that Browser, Language bindings, and their interaction have very large effect sizes with Partial Eta Squared being 0.921, 0.959 and 0.502, respectively. Figure 1 shows box plots for different configurations. Table 17 lists the winning percentages, execution speeds for different configurations, and effect sizes. Again the standard deviations are low giving confidence that our measurements were reliable and resulting in high effect sizes.

**Table 16.** *Test of Between-Subjects effects with dependent variable of memory usage.*

| Source | Type III Sum | df | Mean Square | F | Sig. | Partial Eta Squared |
|---|---|---|---|---|---|---|
| **Corrected Model** | 8660386,20[3] | 19 | 455809,800 | 1083,317 | 0.000 | 0.973 |
| **Intercept** | 1726575310 | 1 | 1726575310 | 4103526,576 | 0.000 | 1.000 |
| **Browser** | 2838851,766 | 3 | 946283,922 | 2249,019 | 0.000 | 0.921 |
| **Tool** | 5606904,197 | 4 | 1401726,049 | 3331,462 | 0.000 | 0.959 |
| **Browser * Tool** | 244053,583 | 12 | 20337,799 | 48,337 | 0.000 | 0.502 |
| **Error** | 241933,562 | 575 | 420,754 | | | |
| **Total** | 1736339618 | 595 | | | | |
| **Corrected Total** | 8902319,766 | 594 | | | | |

Similar to execution time, IE was the best performing browser as can be observed from Table 19, while Chrome was the second best. Firefox had two last place configurations while Opera had three. Table 18 shows that Python and Ruby shared the first place in memory usage while Java was unanimously the one using the most memory. The differences in the percentage behind the first rank are not as big as they were for the execution time because we measured the virtual usage of the entire system and minimum memory requirement for Windows 8.1 (64-bit) system is 2000 megabytes (Microsoft, 2016). The system used about 1340 megabytes or 1,34 gigabytes of virtual memory after startup when idle. Effect sizes are large between the configurations. Overall, Python binding for Selenium in IE was the best configuration while Java binding for Selenium driving Opera was the worst in terms of memory usage.

---

[3] R Squared 0.973 and adjusted R Squared 0.972

**Table 17.** *Mean, median standard deviation, effect sizes and number of accepted results for memory usage. Sorted by mean of memory usage from lowest to highest.*

| Rank | Configuration | Win % | Mean | Median | St. Dev. | Effect measures in comparison to winner | | | n |
|---|---|---|---|---|---|---|---|---|---|
| | | | | | | Cliff's delta | Cohen's d | % behind 1st rank | |
| 1 | Python/IE | 100 | 1521.99 | 1519.50 | 18.431 | - | - | - | 29 |
| 2 | Ruby/IE | 94.7 | 1548.87 | 1544.28 | 20.293 | 0.716 | 1.387 | 1.77 | 30 |
| 3 | Watir/IE | 89.5 | 1570.68 | 1566.17 | 13.573 | 0.977 | 3.008 | 3.19 | 29 |
| 4 | Ruby/Chrome | 78.9 | 1589.17 | 1589.62 | 11.303 | 1.000 | 4.394 | 4.41 | 30 |
| 4 | C#/IE | 78.9 | 1589.73 | 1586.74 | 25.193 | 0.987 | 3.069 | 4.45 | 30 |
| 6 | Watir/Chrome | 73.7 | 1613.05 | 1614.27 | 19.435 | 1.000 | 4.808 | 5.98 | 30 |
| 7 | Python/Chrome | 68.4 | 1630.92 | 1630.75 | 17.916 | 1.000 | 5.984 | 7.16 | 30 |
| 8 | Ruby/Opera | 63.2 | 1654.38 | 1654.29 | 11.383 | 1.000 | 8.643 | 8.69 | 29 |
| 9 | Watir/Opera | 57.9 | 1675.29 | 1675.46 | 12.921 | 1.000 | 9.632 | 10.07 | 29 |
| 10 | C#/Chrome | 47.4 | 1692.23 | 1681.64 | 37.601 | 1.000 | 5.749 | 11.19 | 30 |
| 10 | Python/Firefox | 47.4 | 1694.96 | 1697.19 | 9.110 | 1.000 | 11.898 | 11.36 | 30 |
| 12 | Python/Opera | 42.1 | 1717.23 | 1717.63 | 11.167 | 1.000 | 12.813 | 12.83 | 30 |
| 13 | Ruby/Firefox | 31.6 | 1736.75 | 1739.72 | 16.502 | 1.000 | 12.277 | 14.11 | 30 |
| 13 | Watir/Firefox | 31.6 | 1744.43 | 1742.28 | 12.067 | 1.000 | 14.280 | 14.62 | 30 |
| 15 | C#/Firefox | 26.1 | 1753.47 | 1757.71 | 22.565 | 1.000 | 11.236 | 15.21 | 30 |
| 16 | C#/Opera | 21.1 | 1772.82 | 1774.82 | 29.993 | 1.000 | 10.091 | 16.48 | 30 |
| 17 | Java/IE | 15.8 | 1787.06 | 1783.25 | 23.507 | 1.000 | 12.549 | 17.42 | 30 |
| 18 | Java/Chrome | 10.5 | 1852.42 | 1860.24 | 18.775 | 1.000 | 17.761 | 21.71 | 30 |
| 19 | Java/Firefox | 5.3 | 1952.15 | 1953.56 | 18.408 | 1.000 | 23.353 | 28.26 | 29 |
| 20 | Java/Opera | 0 | 1975.34 | 1974.03 | 31.630 | 1.000 | 17.513 | 29.79 | 30 |

**Table 18.** *Tools/bindings ranked for browsers by their mean for memory usage.*

| Browser | 1st | 2nd | 3rd | 4th | 5th |
|---|---|---|---|---|---|
| Firefox | Python | Ruby | Watir | C# | Java |
| Chrome | Ruby | Watir | Python | C# | Java |
| Opera | Ruby | Watir | Python | C# | Java |
| Internet Explorer | Python | Ruby | Watir | C# | Java |

**Table 19.** *Browsers ranked for bindings/tools by their mean for memory usage.*

| Tool/ Binding | 1st | 2nd | 3rd | 4th |
|---|---|---|---|---|
| C# | Internet Explorer | Chrome | Firefox | Opera |
| Java | Internet Explorer | Chrome | Firefox | Opera |
| Python | Internet Explorer | Chrome | Firefox | Opera |
| Ruby | Internet Explorer | Chrome | Opera | Firefox |
| Watir | Internet Explorer | Chrome | Opera | Firefox |

## 4.3. Calculating length of test scripts

We assume the same amount of test maintenance as we are using the different configuration of the same tool.

As previous research has shown that file size is predictive of maintenance effort (Sjøberg et al., 2013), we computed the file size of the different configurations. Table 20 shows the non-commented lines of code (NCLOC), lines of code (LOC) and file size in kilobytes. Fenton and Bieman (2014) define non-commented lines of code (NCLOC) as a program file listing, with comments and blank lines removed. They add that NCLOC is useful for comparing subsystems, components and implementation languages (Fenton and Bieman, 2014). In addition to blank lines and comments, the imports of libraries were not counted towards NCLOC. Lines of Code are the actual number of lines in the source code as it was run. File size listed in kilobytes contains only the source file containing the code itself, not any other files associated like needed libraries, Webdrivers or generated project files. As can be seen from the Table 20, scripts made for Python bindings had the least NCLOC. Coding conventions for Java (Oracle, 1999) and C# (Microsoft, n.d.) differ in their use of braces, which has an effect on test script lengths. Also the unit testing framework for C# demands the test method to be marked with test method attribute. Some of these differences can be observed by comparing an implementation of a very basic test for different languages and tools, all test methods and functions presented in Figure 2 come from scripts using Firefox as browser.

**Table 20.** *Test script lengths for whole test set used by configurations sorted by calculated lines of code in ascending order*

| Configuration | NCLOC | LOC | File size (kb) |
|---|---|---|---|
| Python/Firefox | 290 | 383 | 19,6 |
| Python/Chrome | 297 | 390 | 19,8 |
| Python/Opera | 303 | 397 | 20,0 |
| Watir/Firefox | 305 | 388 | 14,2 |
| Python/IE | 309 | 402 | 21,2 |
| Watir/Chrome | 310 | 393 | 14,3 |
| Watir/Opera | 311 | 394 | 14,4 |
| Ruby/Firefox | 318 | 402 | 16,6 |
| Ruby/Opera | 326 | 410 | 16,8 |
| Ruby/Chrome | 328 | 410 | 16,8 |
| Java/Firefox | 329 | 425 | 21,4 |
| Java/Chrome | 330 | 426 | 21,4 |
| Watir/IE | 332 | 413 | 15,5 |
| Java/Opera | 340 | 433 | 22,0 |
| Ruby/IE | 340 | 423 | 17,8 |
| Java/IE | 349 | 442 | 22,8 |
| C#/Chrome | 398 | 462 | 25,6 |
| C#/Firefox | 398 | 464 | 25,7 |
| C#/Opera | 398 | 467 | 25,8 |
| C#/IE | 409 | 476 | 27,0 |

```
258   def test_that_verifies_the_url_of_the_statistics_page(self):
259       driver = self.driver
260       driver.get("https://addons.allizom.org/en-US/firefox/addon/firebug/")
261
262       statistics_page = driver.find_element(By.CSS_SELECTOR, '#daily-users > a.stats').click()
263       WebDriverWait(driver, 10).until(EC.title_contains("Statistics"))
264       self.assertIn('/statistics', driver.current_url)
```

A. Test function in Python

```
317   [TestMethod]
318   public void TestThatVerifiesTheUrlOfTheStatisticsPage()
319   {
320       driver.Navigate().GoToUrl("https://addons.allizom.org/en-US/firefox/addon/firebug/");
321       WebDriverWait wait = new WebDriverWait(driver, new TimeSpan(0, 0, 10));
322
323       driver.FindElement(By.CssSelector("#daily-users > a.stats")).Click();
324       wait.Until(ExpectedConditions.TitleContains("Statistics"));
325
326       Assert.IsTrue(driver.Url.Contains("/statistics"));
327   }
```

B. Test method in C#

```
268   def test_that_verifies_the_url_of_the_statistics_page
269       wait = Selenium::WebDriver::Wait.new(:timeout => 10) # seconds
270       $d.get "https://addons.allizom.org/en-US/firefox/addon/firebug/"
271
272       $d.find_element(:css => '#daily-users > a.stats').click
273       wait.until { $d.title.include? "Statistics" }
274
275       assert $d.current_url.include? '/statistics'
276   end
```

C. Test method in Ruby using Selenium

```
260   def test_that_verifies_the_url_of_the_statistics_page
261       $b.goto "https://addons.allizom.org/en-US/firefox/addon/firebug/"
262
263       $b.element(:css => '#daily-users > a.stats').click
264       Watir::Wait.until { $b.title.include? "Statistics" }
265       assert $b.url.include? '/statistics'
266   end
```

D. Test method in Ruby using Watir

```
289   public void testThatVerifiesTheUrlOfTheStatisticsPage() {
290       driver.get("https://addons.allizom.org/en-US/firefox/addon/firebug/");
291       WebDriverWait wait = new WebDriverWait(driver, 10);
292
293       driver.findElement(By.cssSelector("#daily-users > a.stats")).click();
294       wait.until(ExpectedConditions.titleContains("Statistics"));
295
296       assertTrue(driver.getCurrentUrl().contains("/statistics"));
297   }
```

E. Test method in Java

**Figure 2**. Same test method/function with all language bindings and tools.

When comparing the implementations, initializing the wait variable has to be done only when using

Selenium binds with Java, C# and Ruby. Initializing more variables in the setup method would bring the overall NCLOC down, but might not be a good habit in practice, for example timeout settings would be the same for every explicit wait.

Influencing the NCLOC between scripts using the same tool and programming language is the need for additional explicit and implicit waits. This is demonstrated with Figure 3.

As automatic waiting does not work with Webdrivers for Chrome, Opera and Internet Explorer, additional wait statements are needed to be added to the code. First explicit wait (line 67 with Firefox, line 69 with Chrome and line 73 with Internet Explorer in ) is for waiting for a banner like promotional menu, loading on top the page after everything else. Without this explicit wait, the cursor position and focus varies leading to errors when clicking the extensions page link (lines 69, 72 and 76, respectively). Additional implicit wait statement is needed for Chrome and Internet Explorer (lines 70 and 74), because the explicit wait beforehand is fulfilled the moment the promotional menu is detected and execution proceeds to the next line, whereas Webdriver for Firefox waits for promotional menus loading animation. For Internet Explorer, an additional explicit wait is needed (line 77) because execution proceeds immediately to the next line without it, leading to HTML element not found error (as the page has not loaded yet and nothing can be found in its source).

```
64    def test_that_checks_if_the_subscribe_link_exists(self):
65        driver = self.driver
66        driver.get("https://addons.allizom.org/en-US/firefox/")
67        load_promomenu = WebDriverWait(driver, 10).until(EC.visibility_of_element_located((By.XPATH, "//*[@id='promos']")))
68
69        extensions_page = driver.find_element(By.CSS_SELECTOR, "#featured-extensions > h2 > a").click()
70        subscribe_element = driver.find_element(By.CSS_SELECTOR, "#subscribe")
71
72        self.assertIn('Subscribe', subscribe_element.text)
```

A.  Test function in Python for Firefox

```
66    def test_that_checks_if_the_subscribe_link_exists(self):
67        driver = self.driver
68        driver.get("https://addons.allizom.org/en-US/firefox/")
69        load_promomenu = WebDriverWait(driver, 10).until(EC.visibility_of_element_located((By.XPATH, "//*[@id='promos']")
70        time.sleep(2)
71
72        extensions_page = driver.find_element(By.CSS_SELECTOR, "#featured-extensions > h2 > a").click()
73        subscribe_element = driver.find_element(By.CSS_SELECTOR, "#subscribe")
74
75        self.assertIn('Subscribe', subscribe_element.text)
```

B.  Test function in Python for Chrome

```
70    def test_that_checks_if_the_subscribe_link_exists(self):
71        driver = self.driver
72        driver.get("https://addons.allizom.org/en-US/firefox/")
73        load_promomenu = WebDriverWait(driver, 10).until(EC.visibility_of_element_located((By.XPATH, "//*[@id='promos']")))
74        time.sleep(2)
75
76        extensions_page = driver.find_element(By.CSS_SELECTOR, "#featured-extensions > h2 > a").click()
77        load_extensions_page = WebDriverWait(driver, 10).until(EC.visibility_of_element_located((By.XPATH, "//*[@id='subscribe']")))
78        subscribe_element = driver.find_element(By.CSS_SELECTOR, "#subscribe")
79
80        self.assertIn('Subscribe', subscribe_element.text)
```

C.  Test function in Python for Internet Explorer

**Figure 3.** Same test function with three different browsers.

## 4.4.    Errors and faults during test set execution

All run-time errors and faults were logged and are shown in Table 21. Measurements of these executions were not taken into account. The considerably higher amount of runs with errors or faults when using configurations with Internet Explorer is pointing towards it being unstable for testing. Finding Internet Explorer unstable for testing is also supported by the experiences and claims of a blog post by Alister Scott (Scott, 2014). These results are affected by the skill of the programmer and the stability of the system under test. Errors and faults might have been decreased by increasing explicit wait statements. Nevertheless, the high number of errors with IE suggest that its higher performance of execution time comes with the increased risk of producing errors in the test runs.

**Table 21.** *Number of runs with errors and faults for each configuration*

| Rank | Configuration | # of errors | % of errors |
|------|---------------|-------------|-------------|
| 1 | Watir/Firefox | 0 | 0 |
| 1 | Python/Opera | 0 | 0 |
| 1 | Java/Opera | 0 | 0 |
| 4 | Python/Firefox | 1 | 3.22 (1/31) |
| 4 | C#/Chrome | 1 | 3.22 (1/31) |
| 4 | Python/Chrome | 1 | 3.22 (1/31) |
| 4 | Ruby/Chrome | 1 | 3.22 (1/31) |
| 4 | Ruby/Opera | 1 | 3.22 (1/31) |
| 4 | Watir/Opera | 1 | 3.22 (1/31) |
| 4 | Java/Firefox | 1 | 3.22 (1/31) |
| 4 | Java/Chrome | 1 | 3.22 (1/31) |
| 12 | C#/Firefox | 2 | 6.25 (2/32) |
| 12 | C#/Opera | 2 | 6.25 (2/32) |
| 14 | Python/IE | 3 | 9.09 (3/33) |
| 14 | Ruby/IE | 3 | 9.09 (3/33) |
| 16 | Ruby/Firefox | 4 | 11.76(4/34) |
| 17 | Watir/Chrome | 5 | 14.29(5/35) |
| 17 | Java/IE | 5 | 14.29(5/35) |
| 19 | Watir/IE | 13 | 30.23(13/43) |
| 20 | C#/IE | 14 | 31.81(14/44) |

## 4.5.    Combined results

Table 22 assesses how different configurations compare with others across all outcomes by computing rank sum measure. In Table 23 it can be seen that Python is the choice of programming tests for all browsers. This can be mostly attributed to doing good in script length, memory usage and test stability. The question which browser is the best for web-testing, however, has no clear answer. Table 24 shows that all browsers scored victories depending on  the programming language used: IE is the best for C# and Ruby, Firefox is the best for Java, Chrome is the best Python, and Opera is the best when using Watir tool. Overall the position for the top configuration for testing is shared between Python/Chrome and Python/Opera configurations. The worst performance is by C#/Firefox configuration.

**Table 22.** *Configurations combined rank in ascending order*

| Rank | Configuration | Script length | Speed | Memory | Errors | Rank SUM |
|------|---------------|---------------|-------|--------|--------|----------|
| 1 | Python/Chrome | 2 | 10 | 7 | 4 | 23 |
| 1 | Python/Opera | 3 | 7 | 12 | 1 | 23 |
| 3 | Python/IE | 5 | 5 | 1 | 14 | 25 |
| 3 | Python/Firefox | 1 | 10 | 10 | 4 | 25 |
| 5 | Ruby/IE | 14 | 2 | 2 | 14 | 32 |
| 6 | Ruby/Chrome | 10 | 18 | 4 | 4 | 36 |
| 6 | Ruby/Opera | 9 | 15 | 8 | 4 | 36 |
| 6 | Watir/Opera | 7 | 16 | 9 | 4 | 36 |
| 9 | Watir/Firefox | 4 | 19 | 13 | 1 | 37 |
| 10 | Java/Firefox | 11 | 7 | 19 | 4 | 41 |
| 11 | Watir/IE | 13 | 9 | 3 | 19 | 44 |
| 11 | Java/Chrome | 12 | 10 | 18 | 4 | 44 |
| 13 | C#/IE | 20 | 1 | 4 | 20 | 45 |
| 13 | C#/Chrome | 17 | 6 | 10 | 12 | 45 |
| 15 | Watir/Chrome | 6 | 19 | 6 | 17 | 48 |
| 15 | Java/Opera | 14 | 13 | 20 | 1 | 48 |
| 17 | C#/Opera | 17 | 4 | 16 | 12 | 49 |
| 18 | Java/IE | 16 | 3 | 17 | 17 | 53 |
| 19 | Ruby/Firefox | 8 | 17 | 13 | 16 | 54 |
| 20 | C#/Firefox | 17 | 14 | 15 | 12 | 58 |

**Table 23.** *Tools/bindings ranked for browsers by rank sum.*

| Browser | 1st | 2nd | 3rd | 4th | 5th |
|---------|-----|-----|-----|-----|-----|
| Firefox | Python | Watir | Java | Ruby | C# |
| Chrome | Python | Ruby | Java | C# | Watir |
| Opera | Python | Ruby | Watir | Java | C# |
| Internet Explorer | Python | Ruby | Watir | C# | Java |

**Table 24.** *Browsers ranked for bindings/tools*

| Tool/ Binding | 1st | 2nd | 3rd | 4th |
|---|---|---|---|---|
| **C#** | Internet Explorer | Chrome | Opera | Firefox |
| **Java** | Firefox | Chrome | Opera | Internet Explorer |
| **Python** | Chrome | Opera | IE | Firefox |
| **Ruby** | Internet Explorer | Chrome | Opera | Firefox |
| **Watir** | Opera | Firefox | Internet Explorer | Chrome |

## 5. Discussion

Our high level research goal was *to find the most performance efficient configuration for web-testing*. We had four measures of performance: *execution time, memory usage, length of test scripts and stability of the tests*. Next we discuss our results in Sections 5.1-5.6

### 5.1.     The best web-testing configuration when using Selenium & Watir

Section 4.5 presented the results by forming the rank of sums of the four measures. We found that Python programming language was present in all of the top most configurations making it clearly the best language choice for web-testing with Selenium. So whether one decides to use IE, Chrome, Firefox or Opera one should select Selenium with Python language bindings. Furthermore, the choice of single language, i.e. Python, can be argued by the need to verify that web application is working with different browsers (Cross Browser Testing).

With respect to the browser choice no clear winner emerged as each browser had at least one first place depending on the programming language or the tool used, see Table 24.

There seems to be a trade-off between execution speed and stability when using Internet Explorer. If one needs performance, IE is clearly the best in our Windows based test environment. However, choosing IE reduces stability in the test executions. Additionally, extra effort is needed for developing test scripts for Internet Explorer, as more explicit wait statements are needed.

Furthermore, we do not recommend using Opera as the primary choice for web-testing. Currently, the Webdriver needed for Opera has received only four releases in the past 18 months, whereas the browser itself has had 10 major updates during that time. This means that testing has to be done with an older version of the Opera browser or with the increased risk of compatibility issues.

### 5.2.     Summary and Comparison to related work

In evidence based software engineering decisions should not be based on a single study alone. Here we summarize our results and compare them with prior work.

#### 5.2.1. Execution time

We found out that in terms of execution time Selenium is clearly faster than Watir. These results are supported by Angmo and Sharma (2014), graph 3 of their paper on page 734 shows around 14% greater execution time for Watir Webdriver compared with Selenium Webdriver. Angmo and Sharma (2014) do not specify the exact configurations down to version numbers, but in their research Mozilla Firefox is driven by versions of Selenium and Watir that use Webdriver. Our results are in conflict with results by Singh and Tarika (2014), who report over 100% increase in average execution time for Selenium compared with Watir. Differing results are at least in part explained by the use of deprecated Selenium RC by Singh and Tarika (2014) compared with the use of Selenium Webdriver in our research. Configurations used by Singh & Tarika (2014) are reported as Selenium RC driving Mozilla Firefox with test cases written in Java using Selenium IDE plugin, and Watir Webdriver driving Mozilla Firefox with test cases written in Ruby with Testwise Recorder sidebar plugin for Firefox. In our study Watir was on average 50% slower than the fastest Selenium configuration and still on average 4.4% slower than the slowest Selenium configuration.

Recently, Nanz and Furia (2015) performed a study comparing 8 programming languages with respect to the performance of the programs corresponding to the solutions of 745 tasks presented in the Rosetta Code web-site. From the programming languages used in this study, they report programs using Java being the fastest, followed by C#, Python and finally Ruby. Their results correspond with the results of this study by Ruby being the slowest. In contradiction, our study showed that configurations using C# and Python outperformed configurations using Java in execution time. The reason for this contradiction is unknown. The study by Nanz and Furia (2015) measured the performance of different programming languages, while our study measured the performance of the web-testing tools when used with different browsers and programming languages. We think that this contradiction highlights that general purpose performance of programming languages may not be generalizable to situations when programming languages are used as a part of larger configuration.

Another programming language comparison by Prechelt (2000) investigated the differences of seven languages but only two of those languages (Java and Python) were present in our study. Those results show that execution time of Java and Python are nearly identical with Python having a slight advantage.

The browser used influenced execution speed of tests significantly. As can be seen from the results of Wilcoxon tests, highlighted by the winning percentages in Table 13, all three fastest configurations used Internet Explorer.

Table 15 shows the fastest browsers for every configuration, this shows that Internet Explorer is the fastest browser to drive with every tool and binding. Opera is the second fastest and Chrome the third fastest for four of the five tools/bindings. Firefox is the slowest for three of the five tools/bindings.

Widder (2015) compared the performance of various browsers, including Google Chrome (version 45), Internet Explorer 11, Mozilla Firefox 30 and Opera 31. The study compares the performance of browsers with five different benchmarks in total. The most interesting results in the context of this study however, come from the benchmarks measuring performance speed with JavaScript. This is because one of the remaining two benchmarks (Acid3) gives the same grade to every browser tested and the remaining benchmark measures compliance to HTML5 standard instead of pure performance. If only the browsers used in this study are considered, the results of benchmarks used provide an indefinite picture. For example, Internet Explorer performs the best with Sunspider benchmark, but gets the worst result with benchmarks Kraken and Octane 2.0. Counterintuitively, it could be that Internet Explorer and Opera are faster than Chrome and Firefox in our study, as they require a greater number of explicit wait statements in the code. Looking more closely at the trade-offs of automatic waiting versus specific explicit waits could be worthwhile in this regard.

### 5.2.2. Memory usage

In terms of memory usage Java based Selenium configurations used the most memory while depending on the browser Python or Ruby used the least memory. Comparison between Watir and Selenium only shows that performance depends solely on the language binding used in Selenium.

Nanz and Furia (2015) compared the memory use of different programming languages and their results slightly contradicted the results of this study. Their study ranks C# as the most efficient programming language from the languages used in this study, followed by Java and finally by Ruby & Python in a tie. If the results of this study for C# and Java are disregarded, the results of Nanz and Furia (2015) are much closer to the findings of this study. Systems memory use when using configurations with Ruby (both Selenium and Watir) or Python were close to each other, as suggested by Nanz and Furia (2015). In terms of percentage, the memory use of the whole system when using Python/Opera is 3.8% higher than when using Ruby/Opera, this being the highest difference between two configurations using the same browser when C# and Java are disregarded.

Study by Prechelt (2000) showed that in average Java programs require over 100% more memory than Python programs. Our results would be similar, if we chose to exclude the base memory required by the system. The memory exclusion can be done simply by subtracting 1340 from the memory consumption figures of Table 17.

### 5.2.3. Script length

Non commented lines of code are shown in Table 20. The length of test scripts is influenced by the language bindings together with the browser used. For every configuration using the same browser the language bindings from using least lines of code to most lines of code were: Python bindings, Watir, Ruby bindings for Selenium, Java bindings for Selenium and finally C# bindings. C# bindings used roughly 35-40% more lines than Python depending on the browser used. With more extensive test methods the difference in line count would not be as big, as numerous additional code lines are braces and tags used by the unit testing framework. For every language binding, the configuration using Firefox used the least amount of lines and configuration using Internet Explorer the most. This is due to the additional explicit wait statements needed for configurations using other browsers than Firefox.

The results regarding script length are backed by Nanz & Furia (2015). In their paper Python uses the least amount of lines, followed in order by Ruby, Java and C#. They report a clear division between more dynamically typed functional languages (Python & Ruby) and more statically typed objected oriented languages (Java & C#). This division is not clearly seen in the results of this study, as the results for length in non-commented lines of code for scripts using Java is quite close to scripts done in Python and Ruby. For example scripts using Java were only 2% - 3.5% longer than scripts using Ruby with Selenium bindings. Difference was the biggest solely between C# and other language bindings.

### 5.2.4. Test stability

We found that IE was the most unstable browser for testing. The choice of programming language did not seem to affect test stability. Also a blog post by Alister Scott (Scott, 2014) found that Internet Explorer is unstable for testing although no quantitative evidence is provided by the source. Test stability is an important industrial problem that merits further investigation. In future we plan to significantly increase the test executions to get improved evidence on the test stability.

## 5.3.    Future Research

Performance of the test tool in test execution is not the only factor to be taken into account when choosing the testing tool. In this section, we present other factors affecting web testing that could be included in future studies of this topic.

### 5.3.1. More factors – Maintenance, manual effort of test creation and more holistic approaches

Future studies could assess tools with wider range of factors than we do in this study. Here we outline three such factors but the list is by no means comprehensive.

Maintenance of test scripts is heavily dependent on the way the test scripts are implemented. For better maintainability, software patterns such as Page Object and Factory can be utilized. Previous work by van Deursen (2015) and Leotta et al. (2016) focus on these patterns in web testing context. Investigating how these patterns work with different bindings of Selenium could be a worthwhile effort. Additionally, measuring how long fixing broken test scripts takes on each tool could be a fruitful direction of research.

Lines of code has been used as a measurement of used effort for test code implementation in scientific literature (e.g. Kamei et al. 2010), but it would be useful to gather such metrics as time spent on implementing the tests as well. Originally time implementing test scripts for this study was measured. However, the tools compared in this study are quite similar to use, hence learning and familiarity with the programming language affects how fast the tests are implemented for each tool. In practice, the most effort was spent on tests implemented first and effort spent lessened incrementally afterwards due to learning. This was the case even when changing the programming language. In our experience, differentiating the experience of the usage of the tools and the language bindings is troublesome, as most of the API function calls for different language bindings are equivalent.

We are aware of studies by Alegroth et al.(2016) in the area of Visual GUI testing that address such tools like Sikuli, in industrial context with holistic case studies. Providing more holistic comparison, e.g. also including interviews with practitioners and using particular cases as bases would be important to have a better view how Selenium compares with other tools.

### 5.3.2. More tools

Many other less popular tools for web-testing exist such as Ranorex, JAutomate, Sikuli and Selenium IDE (Raulamo-Jurvanen et al., 2016; Yehezkel, 2016). Our test set was not implemented into every tool, because of limited amount of resources. Adding more tools to the benchmark is something we look for in the future. Furthermore, we believe that examining different versions of Selenium is specifically meaningful, as it is currently the most popular testing tool in the industry (Raulamo-Jurvanen et al., 2016; Yehezkel, 2016). Specifically, in the industrial survey conducted by Raulamo-Jurvanen et al. (2016) Selenium ranked significantly higher than any other web automation tool (table 3). Selenium was the second most used tool in overall ranking behind only programming language Python, while no other desktop based web automation tool ranked in the top ten.

Comparing web testing tools based on their features could also be a worthwhile task, e.g. features such as taking screenshots when tests fail, can be highly beneficial for analyzing test results. However, when comparing tools it is noteworthy that some tools are designed for a specific purpose (Selenium), while others are part of a large family of tools (Ranorex). Hence taking into consideration the context of use is also important.

Plenty of further research in the area of test automation benchmarking could be done. One of our future goals is to provide our benchmarking scripts in machine images as such, for example to Amazon EC2. Currently all of our scripts and measurement results are available in BitBucket (https:goo.gl/gEqX07), but running the scripts requires maintenance updates.

Adding mobile environments would also be an interesting extension. This research did not touch Webdrivers for mobile context like Selendroid (2015), Appium (2016) or Windows Phone (Microsoft, 2013). For desktop context, there are also Webdrivers for Microsoft Edge (Microsoft, 2015) and Apple's Safari (SeleniumHQ., 2016b).

There is also a number of popular tools and frameworks that can be integrated with Selenium and Watir but whose performance is unknown. Build automation tools Maven and Jenkins can be used together with Selenium and Watir, but no scientific literature about the subject was found when making the literature review for this study. Reports on the problems with integrating tests using Webdriver and Jenkins exist (SauceLabs, n.d.); the origin, fixing or bypassing of these problems might be a meaningful direction for new research.

### 5.3.3. More systems under test

Our tests were implemented for one system under test, Mozilla Addons store. The generalizability of the results would increase by adding tests to different systems. However, we think our tests study offers a reasonable starting point, as many complex tasks such as navigating dynamic menus, taking into account Ajax content, logging into the system and saving user generated content to the database are performed. Such features are highly typical in any web-application.

## 5.4. Benefits of faster test execution

Several sources state that performance such as speed or stability of testing tools matter as already argued in Section 1. We can find roughly two types of claims in the this area.

The first is related to the soft factor benefits of faster test executions as part of build and integration process. An industry report by Rogers (2004) claims that long builds (30-40 minutes) cause infrequent integration which in turn increase integration effort while short builds (2 minutes or less) lead to situation where integration is done at will. A similar point is also made by (Herzig et al., 2015) who (after providing thorough calculations of the money-wise benefits of faster test execution cycles) states that "the actual values are secondary, it is important that the achieved productivity increases through faster integrations". Rapid feedback is seen as a core principle behind continuous integration (Fowler, 2006) and is considered highly important for productivity also in each developers' machine (McIver, 2016) . It appears that modelling the benefits of faster feedback in terms of money is difficult. In our past work, we performed a brief interdisciplinary review of build waiting time (Mäntylä and Laukkanen, 2015) and found that various cognitive effects, for example loss of attention and errors, or emotional effects, e.g. anger and anxiety can results from increased waiting time. Importantly, the flow of development is also disturbed while waiting for the build to finish. The flow state is defined as having characteristics such as intense focused concentration to present moment, loss of or reflective self-consciousness, and experiencing of the activity as intrinsically rewarding (Nakamura and Csikszentmihalyi, 2014). Hence with faster test execution times for automated tests, the disturbance to the flow of the developers decreases. We hope that future studies can offer money-wise computations with respect to the speed of feedback and importance of flow in software development.

Second, there are studies trying to assess the money-wise of testing that mainly ignore the topics mentioned in the above paragraph but offer additional viewpoints. Deissenboeck et al. (2008) highlight that "the execution costs of the tools can be divided into (1) tool execution costs and (2) report analysis costs". However, the authors falsely claim that the tool execution costs are negligible because the execution does not need human attention. A study at Microsoft has shown the savings of 1.5 million dollars for execution time cost spent in testing whereas report analysis costs measured with the investigation of false reports were only 60 thousand dollars during the same period of Microsoft Windows testing (Herzig et al., 2015). On the other hand, the Microsoft study also showed other cases like the Dynamics product where only 20 thousand dollars could be saved through test execution time and 2.3 million could be saved through report analysis.

For illustrative purposes only, we made similar calculations for Mozilla Add-ons store as was done in (Herzig et al., 2015). During its life time (almost exactly 7 years) Mozilla Add-ons store has had 27,175 commits (date 26 Oct 2016). Currently, it has 160 tests. If we assume that for each commit all the tests are executed as is typically done in continuous integration environment, we come up with 2,174,000 test case execution (80*27,175) (We assume linear evolution in the number of tests in the project thus on average we have 80 test execution for each commit). We use our test executions times from Table 14 divided by our number of tests as an average test case execution time. We use the price for computing power as (Herzig et al., 2015) $0.03 $/min multiplied by two as we are testing server client application. We use our failure rate percentage from Table 21. We use the same average test inspection cost of false alarm $9.60 per inspection as (Herzig et al., 2015). We find that majority of the cost savings comes from preventing false alarms. If one can reduce false alarms from 3.2% to 0% the false alarms results in the cost savings of $672,027 while the savings in the test execution time (even between the fastest and the slowest) is only negligible $8000.

## 5.5. Configuration choice popularity in the Web

Having performed the benchmark experiments, we became curious to investigate whether the best choices are already adapted in the industry. Web-scraping methodology gives a lightweight proxy to the popularity of testing tools (Raulamo-Jurvanen et al., 2016). The reasoning is that the number of web-hits from various forums reflects the popularity of a particular tool. Thus, we are interested in finding whether the best configurations used were also the ones receiving most mentions in terms of various web-statistics. The results are in Table 25.

**Table 25.** *Web-statistics & rankings for the configurations.*

| Binding | | | GoogleHits | GoogleHits | StackOverflow | | | Total | |
|---|---|---|---|---|---|---|---|---|---|
| **Tool** | **Language** | **Browser** | **GH_1** | **GH_2** | **Qs** | **ViewCount** | **VC/Qs** | **Sum** | **Rank** |
| **Selenium** | Java | Firefox | 415000 [2] | 172000 [2] | 238 [1] | 15984 [1] | 67,2 [8] | 14 | 1 |
| **Selenium** | Java | Chrome | 402000 [3] | 135000 [3] | 175 [2] | 12670 [2] | 72,4 [6] | 16 | 2 |
| **Selenium** | Python | Firefox | 241000 [6] | 114000 [4] | 78 [3] | 4966 [3] | 63,7 [11] | 27 | 3 |
| **Selenium** | Python | Chrome | 216000 [8] | 88200 [5] | 63 [4] | 4131 [4] | 65,6 [9] | 30 | 4 |
| **Selenium** | Ruby | Chrome | 228000 [7] | 67000 [11] | 17 [8] | 1933 [7] | 113,7 [1] | 34 | 5 |
| **Selenium** | Java | IE | 118000 [12] | 68000 [9] | 20 [6] | 2042 [6] | 102,1 [2] | 35 | 6 |
| **Selenium** | C# | Chrome | 180000 [9] | 69100 [8] | 10 [9] | 907 [9] | 90,7 [3] | 38 | 7 |
| **Selenium** | Java | Opera | 114000 [13] | 45000 [12] | 41 [5] | 3500 [5] | 85,4 [5] | 40 | 8 |
| **Selenium** | Ruby | Firefox | 174000 [10] | 75400 [6] | 20 [6] | 1261 [8] | 63,1 [12] | 42 | 9 |
| **Selenium** | Ruby | Opera | 1210000 [1] | 227000 [1] | 3 [14] | 129 [14] | 43,0 [17] | 47 | 10 |
| **Selenium** | C# | Firefox | 144000 [11] | 72700 [7] | 6 [11] | 383 [12] | 63,8 [10] | 51 | 11 |
| **Watir** | Ruby | Firefox | 26100 [18] | 15700 [16] | 5 [12] | 440 [10] | 88,0 [4] | 60 | 12 |
| **Selenium** | Python | Opera | 69200 [14] | 19300 [15] | 8 [10] | 418 [11] | 52,3 [14] | 64 | 13 |
| **Selenium** | Ruby | IE | 270000 [5] | 67900 [10] | 1 [16] | 51 [18] | 51,0 [16] | 65 | 14 |
| **Selenium** | Python | IE | 289000 [4] | 26900 [13] | 1 [16] | 28 [19] | 28,0 [19] | 71 | 15 |
| **Watir** | Ruby | Opera | 37500 [17] | 14100 [18] | 1 [16] | 68 [16] | 68,0 [7] | 74 | 16 |
| **Watir** | Ruby | Chrome | 20100 [19] | 12500 [19] | 5 [12] | 308 [13] | 61,6 [13] | 76 | 17 |
| **Selenium** | C# | IE | 52600 [15] | 25900 [14] | 1 [16] | 52 [17] | 52,0 [15] | 77 | 18 |
| **Watir** | Ruby | IE | 10200 [20] | 5060 [20] | 2 [15] | 84 [15] | 42,0 [18] | 88 | 19 |
| **Selenium** | C# | Opera | 42400 [16] | 14500 [17] | 0 [20] | 0 [20] | 0,0 [20] | 93 | 20 |

The Google hits were searched manually search string "<tool> <language> <browser>" (GH_1 in the table). "Selenium Ruby Opera" was the combination with highest number of Google hits, 1210000. It had nearly three times as many hits as the combination having the second most number of Google hits, "Selenium Java Firefox", with 415000 hits. The combination "Selenium Java Chrome" had 402000 Google hits while the combination having the next highest Google hits, "Selenium Python IE" was significantly lower with 289000 hits.

When adding the word "webdriver" to the previous search string the rankings of the combinations changed slightly (GH_2 in the table). The first three combinations with the highest count of Google hits remained the same (with hit counts as 227000, 172000 and 135000, respectively). Having the word "webdriver" in the search string reduced the number of hits radically and at the same time, made the differences smaller than with the previous search string.

Finding the number of StackOverflow questions from StackExchange (www.stackoverflow-com) for our configurations was tricky. There are several different types of tags for those tools: selenium (29k), selenium-webdriver (15,7k), selenium-ide (1,5k), selenium-rc (1,4k), selenium-chromedriver (1,2k), selenium-grid (775), selenium-firefoxdriver (218), rselenium (122), selenium-grid2 (88), selenium-server (39), selenium-fitnesse-bridge (19), selenium2library (14), selenium-builder (9), selenium-ruby (1), flash-selenium (1), selenium2 (15,7k), selenium-webdriver-c# (15,7k) and selenium-webdriver-java (15,7k). Therefore, we decided to use a simple approach: to find questions having the words "<tool> <language> <browser>" appearing in the body of the question. That way the search would be also in line with the searches for the Google hits. The time period used for fetching the data for was set to few months 1.1.2016-30.4.2016.

The results indicated that Selenium and Java with Firefox and Chrome browsers were the combinations with most questions, 238 and 175, respectively. Combinations of Selenium and Python with Firefox and Chrome browsers were the combinations having the next most questions, 78 and 63, respectively. Combinations with Watir and Ruby & Selenium and C# in combination with Opera and IE browsers seemed to have the least questions in StackOverflow. The combination of Selenium, C# and Opera had no questions at all in the search.

For the view counts the questions were in the same order as for the actual number of questions. The number of view counts seemed to be rather in line with the number of questions. The questions for the combinations of Selenium, Java & Firefox and Selenium, Java and Chrome had 15984 and 12670 view counts, respectively, while the next combination Selenium, Python and Firefox had only nearly 5000 views. When looking at the average number of view counts for the questions, the combination of Selenium, Ruby and Chrome has the highest value, almost 114 views per questions.

When computing a rank sum of our web-scraping figures, the combination with the highest rank seems to be Selenium, Java and Firefox, followed by Selenium Java and Chrome, Selenium, Python and Firefox and Selenium Python and Chrome. Watir, in general and combinations of Ruby, C# and Opera and IE seem to populate the last places of the ranking.

For such combinations of a tool, programming language and browser finding relevant tweets is challenging. Thus, we decided to search for all tweets for Selenium and Watir for the time period between the 1st of January and the 30th of April in 2016. Selenium had clearly more tweets than Watir, 20247 vs. 3526, respectively. In those 20247 tweets for Selenium Java was mentioned 3312, Python 1862, Ruby 487 and C# 473 times. This is in line with other web-scraping results. In those tweets for Selenium the browsers were mentioned as follows, Firefox 433, Chrome 347, IE 27 (as "IE" (18) and "internet explorer" (9)) and Opera 21 times. Again, this is in line with other web-scarping sources.

Overall Java was the most popular language followed by Python in Web-scraping. In our experiment, Python performed the best overall. Java's performance varied between the 3rd and the 5th place. Thus, it appears that our experiment results are partially implicitly known by the industry as Python did fairly well. On the other hand the popular choice of using Java with Selenium is not supported by our results. Overall the best configuration in our experiments was Selenium, Python, Chrome that was ranked 4th in web scraping. Putting the results other way around the most popular web scraping configuration was Selenium, Java, Firefox that was ranked as 10th in our experiments. The Table 26 and Table 27 provide rankings for tools/bindings and browsers based on data from Web-scraping in comparison to results to the study, see Table 23 and Table 24.

Popularity of programming languages may influence our web-scraping results, for example at the time of writing this article searching for tags in StackOverflow demonstrates differences. Java and C# tags have been used around 1,15 and 1 million times, respectively, whereas Python and Ruby have only around 640k and 170k questions with their tags, respectively. Popularity of Selenium can be demonstrated this way as well, Selenium tag has been used around 33,5k times in StackOverflow, while the same number is 1977 for Watir, none for JAutomate and 519 for Sikuli at the time of writing this article.

**Table 26.** *Tools/bindings ranked for browsers by rank sum (web-scraping).*

| Browser | 1st | 2nd | 3rd | 4th | 5th |
|---|---|---|---|---|---|
| **Firefox** | Java | Python | Ruby | C# | Watir |
| **Chrome** | Java | Python | Ruby | C# | Watir |
| **Opera** | Java | Ruby | Python | Watir | C# |
| **Internet Explorer** | Java | Ruby | Python | C# | Watir |

**Table 27.** *Browsers ranked for bindings/tools(web-scraping).*

| Tool/ Binding | 1st | 2nd | 3rd | 4th |
|---|---|---|---|---|
| **C#** | Chrome | Firefox | Internet Explorer | Opera |
| **Java** | Firefox | Chrome | Internet Explorer | Opera |
| **Python** | Firefox | Chrome | Opera | Internet Explorer |
| **Ruby** | Chrome | Firefox | Opera | Internet Explorer |
| **Watir** | Firefox | Opera | Chrome | Internet Explorer |

## 5.6. Limitations

Here we outline the limitations of our results. The purpose is given an honest account of the research. The aim of this section is to inform other researchers how they can do even better based on our experiences.

Results of this study cannot be used to make judgments on the efficiency of defect detection capabilities of these tools. However, these tools are usually used together with libraries providing different assert statements, leaving much responsibility to the test developer and test case designer.

Limiting the generalizability of the results is the fact that tests were run on a home desktop computer that was not what we would call a fully standardized environment. We have documented the test environment to our best ability (see Tables 5-7). Another limiting factor is running tests for a single web application. Adding more tests (to other services or applications) to benchmark test set, the generalizability would increase.

Measuring virtual memory usage purely programmatically without the human element of reaction time would make the results more reliable and robust. However, as our standard deviations of the memory usage between test runs are very low, see Figure 1, we suspect that we would experience no major changes in our results. Furthermore, the very high partial Eta squared values for memory consumption (0.973), suggest that external error source of variation was not present. It seems likely that the deviations between test runs could have become smaller but that the order between configurations would not have changed had we used a purely programmed approach.

Assert statements used in different implementations of the test set vary in name, because the basic libraries used by different programming languages have different assert methods. The equivalent use of assert statements was strived for, but in principle these could affect the performance of different configurations.

The tests that were re-programmed for the purposes of this study were originally made by community developers. These developers may have extra skills and gimmicks developing the tests for Python bindings that may have transferred partly to re-programmed test scripts. Original tests were also designed for Python bindings. Thus, there is a danger that our results favour Python programming language and we welcome further studies of this topic. Our test scripts as well as our measurements are available for anyone to investigate at a repository (http:goo.gl/gEqX07).

Outcomes of the test runs may not mirror outcomes in different contexts, for example the code can be run from integrated development environment or remotely using Selenium remote. Context of use may thus influence the performance of the testing configuration.

The server hosting the web application under test was not under any control of the authors of this study, thus server load and capability which could alter the results are not known. However, this was combated by running the tests in the series of ten and combined to samples of thirty. Finally, the very high partial Eta squared (0.994) measured from ANOVA, suggest that nearly all of the variation was due to our configurations and not due to external factors.

## 6. Conclusions

To our knowledge, this paper is one of the first reports that provide public empirical evidence on the performance of web-testing automation tools. The prior work in this area is limited as demonstrated in Section 2.

We demonstrate big differences in performance efficiency between web-testing configurations using Watir and Selenium with different programming languages and browsers. The slowest configuration was 91% slower than the fastest while the most memory hungry configuration used 30% more memory than the one with the lowest memory consumption. The effect sizes between the extremes were also very large for both execution time (Cohen's d=41.5, Cliff's delta=1.0) and memory usage (Cohen's d = 17.5, Cliff's delta=1.0).

Selenium tool with Python bindings was found as the best choice for all browsers. The single best configuration was Selenium with Python bindings on Chrome browser. It tied the first place with Selenium with Python on Opera browser, see Section 4.5. However, we consider this as a secondary option as Selenium updates to Opera browser are offered less frequently than for Chrome. We did our measurements in Windows environment and performance wise (memory, and execution time) IE was the best browser. However, test executions in IE resulted in poor test stability with several abnormal test terminations. Thus, in IE the number of flaky tests will be higher than with other browsers and this will increase test result analysis effort.